# Constraining prebiotic chemistry through a better understanding of Earth's earliest environments

**Timothy W. Lyons** (University of California, Riverside), Phone: 951-827-3106, Email: timothy.lyons@ucr.edu

Co-authors: **Karyn Rogers** (Rensselaer Polytechnic Institute), **Ramanarayanan Krishnamurthy** (Scripps Research Institute), **Loren Williams** (Georgia Institute of Technology), **Simone Marchi** (Southwest Research Institute), **Edward Schwieterman** (Univ. of California, Riverside), **Dustin Trail** (Univ. of Rochester), **Noah Planavsky** (Yale Univ.), **Christopher Reinhard** (Georgia Institute of Technology)

Co-signatories: **Vladimir Airapetian** (NASA GSFC/American Univ.), **Jan Amend** (Univ. of Southern California), **Ariel Anbar** (Arizona State Univ.), **Jose Aponte** (NASA GSFC/Catholic Univ. of America), **Giada Arney** (NASA GSFC), **Laura Barge** (NASA Jet Propulsion Laboratory, California Institute of Technology), **Steven Benner** (Foundation for Applied Molecular Evolution), **Roy Black** (Univ. of Washington), **Dennis Bong** (Ohio State Univ.), **Donald Burke-Aguero** (Univ. of Missouri), **Jeffrey Catalano** (Washington Univ.), **David Catling** (Univ. of Washington), **George Cody** (Carnegie Institution for Science), **Jason Dworkin** (NASA GSFC), **Jamie Elsila** (NASA GSFC), **Paul Falkowski** (Rutgers Univ.), **Daniel Glavin** (NASA GSFC), **Heather Graham** (NASA GSFC/Catholic Univ. of America), **Nicholas Hud** (Georgia Institute of Technology), **James Kasting** (Penn State Univ.), **Sarah Keller** (Univ. of Washington), **Jun Korenaga** (Yale Univ.), **Ravi Kumar Kopparapu** (NASA GSFC), **Susan Lang** (Univ. of South Carolina), **Michaela Leung** (Univ. of California, Riverside), **Tom McCollom** (Univ. of Colorado, Boulder), **Victoria Meadows** (Univ. of Washington), **Ulrich Muller** (Univ. of California, San Diego), **Vikas Nanda** (Rutgers Univ.), **Stephanie Olson** (Purdue Univ.), **Niki Parenteau** (NASA Ames Research Center), **Ulysse Pedreira-Segade** (Rensselaer Polytechnic Institute), **Alexandra Pontefract** (Georgetown Univ.), **Milena Popovic** (Blue Marble Space Institute of Science), **Richard Remsing** (Rutgers Univ.), **Scott Sandford** (NASA Ames Research Center), **Laura Schaefer** (Stanford Univ.), **Morgan Schaller** (Rensselaer Polytechnic Institute), **Jacob Shelley** (Rensselaer Polytechnic Institute), **Everett L. Shock** (Arizona State Univ.), **Danielle Simkus** (NASA GSFC), **Rachel Smith** (NC Museum of Natural History/Appalachian State Univ.), **Andrew Steele** (Carnegie Institution for Science), **Geronimo Villanueva** (NASA GSFC), **Bruce Watson** (Rensselaer Polytechnic Institute)

## 1. MOTIVATIONS: A SUMMARY

Any search for present or past life beyond Earth should consider the initial processes and related environmental controls that might have led to its start. As on Earth, such an understanding lies well beyond how simple organic molecules become the more complex biomolecules of life, because it must also include the key environmental factors that permitted, modulated, and most critically facilitated the prebiotic pathways to life's emergence. Moreover, we ask how habitability, defined in part by the presence of liquid water, was sustained so that life could persist and evolve to the point of shaping its own environment. Researchers have successfully explored many chapters of Earth's coevolving environments and biosphere spanning the last few billion years through lenses of sophisticated analytical and computational techniques, and the findings have profoundly impacted the search for life beyond Earth (e.g., Schwieterman et al., 2018). Yet life's very beginnings during the first hundreds of millions of years of our planet's history remain largely unknown—despite decades of research. This report centers on one key point: that the earliest steps on the path to life's emergence on Earth were tied intimately to the evolving chemical and physical conditions of our earliest environments. Yet, a rigorous, interdisciplinary understanding of that relationship has not been explored adequately and once better understood will inform our search for life beyond Earth. In this way, studies of the emergence of life must become a truly interdisciplinary effort, requiring a mix that expands the traditional platform of prebiotic chemistry to include geochemists, atmospheric chemists, geologists and geophysicists, astronomers, mission scientists and engineers, and astrobiologists.

## 2. THE CONTEXT

In short, unraveling how and when life began on Earth, as for any system we might be exploring, requires knowledge about the early environmental backdrop of those advances. Historically, studies of the chemical origins of life start with the hypothesis that mixtures of simple small molecules under the influence of various energy sources and early Earth environments would lead to the building blocks of life (amino acids and other biomolecules) and that interactions among these molecules eventually lead to life. However, while experiments have synthesized some of the building blocks (e.g., amino acids), it remains a first-order challenge to demonstrate the spontaneous synthesis of many other essential components of extant life, such as the chemical building blocks of RNA (nucleotides) and cell membranes (phospholipids), particularly under conditions that faithfully represent realistic early Earth environments. Several biochemical precursors—such as amino acids, nucleobases (e.g., adenine), short chain fatty acids, metabolites, and sugars—have been identified in meteorites, supporting their likely availability on early Earth. However, even if the availability of biochemical precursors is granted, there still has been no experimental demonstration nor consensus view on how these units could come together to eventually give rise to the molecular units needed for life.

This conundrum has generated interest in alternative chemical mechanisms that may have been successful in generating many of the needed target molecules, such as the purine and pyrimidine nucleotides of RNA (e.g., Powner et al., 2009). However, many of these experiments involve complex and highly prescribed networks of sequential chemical steps such that their likelihood on early Earth has been questioned. This gap, in turn, has spurred searches for alternative chemistries beyond the DNA/RNA/protein central dogma, such as pre-RNA coevolution of multiple polymer types or hydrothermal vent metabolism scenarios, which may be more compatible with early Earth criteria. Broadly speaking, though, none of the simple or complex biochemistries considered to date have been demonstrated to function under geologically realistic conditions. Thus, experimental syntheses of biomolecules under assumed early Earth conditions have not gained universal acceptance, while the geological, geophysical, and geochemical details of the early oceans, atmosphere, and crust remain enigmatic. Nonetheless, there is an inevitable coevolution of early Earth environments and the prebiotic pathway that gave rise to life, and understanding this relationship will require unprecedented cooperation among the diverse disciplines that investigate these systems.

Such interaction and the essential back-and-forth among very different communities has been lacking in part because long-standing disagreements within each group have led to insularity. We contend that these long-standing roadblocks can fade with better integration of science across wide-ranging communities, including geologists and prebiotic chemists, so that related constraints on possible boundary conditions can be explored both independently and in light of what other groups are predicting. It is these conversations that could help define the frontiers of research by identifying, for example, (i) the most critical unknowns of early Earth environments relative to prebiotic chemistry, (ii) the chemical/physical necessities shared among the prebiotic chemical models and experiments, and (iii) the prebiotic chemistry scenarios that likely lie outside of planetary reality. Missing more generally is the full range of real-time conversations and collaborations required to explore life's beginnings in a rigorous environmental context. Critically, we now recognize the limitations of a business-as-usual approach and are poised to make the most of the vast opportunities afforded by a more interdisciplinary strategy.

## 3. BUILDING BRIDGES AND PRIORITIZING

Two steps must be taken to bring us closer to how the processes of life began on Earth and might have elsewhere in the universe. The first is the deep and broad integration of the communities outlined above. The second is a larger research investment in studies of Earth's environments at their beginnings, including emphasis on parameters key to understanding life's start. Without that context we have little hope of resolving what prebiotic chemical scenarios were possible. Such work now is often performed in a paleoenvironmental vacuum. What follows are specific suggestions for environmental parameters that are little understood but are central to most if not all prebiotic models for life's beginnings.

### 3.1 Water, water everywhere

The record of old rocks (from greater than 3.8 to 4.0 billion years ago or Ga) ranges from extremely patchy to nonexistent for our deepest history, forcing us to rely on the remarkable historical archive found in well-preserved zircons with ages that approach 4.4 Ga. Geochemical data from among the oldest of those sand-sized mineral grains suggest early recycling of seafloor that was previously altered in the presence of liquid water at Earth's surface—providing surprising and convincing evidence for very early oceans, cool surface temperatures, early crustal differentiation, and even incipient plate boundary interactions possibly including subduction (Harrison, 2020). As a major proof of concept, arguments for very early oceans, in contrast to prior consensus views of a 'hellish' Hadean world (from the birth of the planet to 4.0 Ga), have shifted the conversation away from magma oceans and high surface temperatures to a world that could have permitted and even favored the beginnings and sustained evolution of early life under persistently habitable conditions—even during Earth's first few hundreds of millions of years.

### 3.2 Land, ho?

The hypothesized Moon-forming event likely marks the initiation of a broad, though not necessarily monotonic, cooling towards a habitable planet. By this time, probably in the first 100 million years, it is likely that Earth's metallic core and bulk silicate mantle had already undergone chemical and physical separation. However—there is no uniform consensus about how and when the Earth generated crustal material analogous to what we know as continental crust today (that is, silica-rich, Fe-poor buoyant rock) following generation of a primordial crust rich in Fe and Mg but poor in Si. Several unresolved issues are pertinent to this debate including the volume of buoyant crust generated versus time (Figure 1); the tectonic regimes operating on Earth; the distribution of heat-producing elements between the crust, mantle, and core and related consequences for surface environments; and the thermal state of the mantle, which directly impacts the buoyancy (and recycling) of crust (Dhuime et al., 2012; Korenaga, 2018). Crustal recycling could, for example, facilitate nutrient recycling in the oceans.

Given broad agreement that there was liquid water in large volumes interacting with the crust before 4.0 Ga (but with only sparse evidence before 4.3 Ga), one critical question that remains is the timing of the emergence of subaerial crust. Whether the crust is emergent or not reflects the interplay between the thermal state of the mantle and the tectonics of the planet balanced against the volume of liquid water at the surface. Fundamental questions remain about the timing of initial continent formation (Figure 1) and the likelihood of its emergence above the surrounding seas (Korenaga, 2018). The importance of exposed land (e.g., Benner et al., 2020)—including related wet-dry cycles that could drive the assembly of the building blocks and their transition to self-sustaining functional systems (Damer & Deamer, 2020) and concentrate compounds through evaporation (e.g., Toner and Catling, 2019)—is central to many views of the prebiotic world, yet we know almost nothing about the details, likelihood, and timing of early land masses.

Beyond the global view, we need to understand solid Earth and volatile interactions at the scale of local early environments. An emergent landmass is one possibility for an important local environment. Information about local settings may be within reach. Tang et al. (2019), for example, found evidence for pre-3.9 Ga chlorine in zircons, suggesting hydrothermal brines interacted with the lithosphere inside the nascent crust. Most of the direct evidence that has shaped our understanding and invigorated debate about the pre-4.0 Ga Earth has come from detrital zircons from the Jack Hills, Australia (Harrison, 2020). Many of the essential environmental questions are conceivably addressable through study of this ancient resource, particularly when bridged to experimental work and geophysical modeling. In addition to this now-famous location, there are 14 other sites with pre-4.0 Ga zircons (Harrison, 2020). With the exception of the 4.03 Ga Acasta Gneiss, current work on these samples lags behind their potential, offering the possibility of entirely new windows into the earliest Earth. Finally, it is even possible that some of the earliest terrestrial fragments may be ultimately recovered as meteorites on the Moon. These putative fragments of Earth on the Moon could be relatively pristine because they escaped a ~four-billion-year residence on a geodynamically active Earth. The recent reinvigoration of lunar research, with the possibility of sample return in the near future, warrants careful attention for those caring about Earth's earliest environments.

### 3.3 The impact of impacts

The Earth formed by the accretion of Mars-sized embryos and countless smaller planetesimals (e.g., Morbidelli et al., 2012). The Moon-forming impact (~4.5 Ga; Barboni et al., 2017) is traditionally considered the last major accretionary event. What is often underestimated, however, is that the young Solar System was still evolving rapidly in the aftermath of Moon formation, with major implications for the earliest evolution of our planet. Collisional models and geochemical data constrain the range of possible terrestrial bombardment histories, as well as likely impactor sizes and compositions (e.g., Marchi et al., 2014; Figure 2). This history of bombardment delivered the final 0.5-2.5% of Earth's mass within 0.5-1 Gyr after the formation of the Moon (Walker, 2009; Marchi et al., 2018). If spread evenly across Earth's surface, this addition would comprise a 10-50 km-thick layer, illustrating the important global consequences of bombardment for near-surface environments.

As prebiotic chemical processes that led to life were taking place, the Hadean Earth was subject to a tremendous and ongoing bombardment by leftover planetesimals that fundamentally influenced its near-surface environments and modulated its chemical and geophysical evolution. Early collisions have been invoked both as potential inhibitors and contributors to the prebiotic chemistry and conditions that led to life. For example, impacts have been associated with havoc in near-surface environments (Sleep et al., 1989) but also the delivery of key prebiotic compounds such as amino acids, sugars, purine, and reactive phosphorus (Furukawa et al., 2019; Callahan et al., 2011) and possible creation of a favorable highly reducing atmosphere (Zahnle et al., 2020). The full gamut of impact-induced processes is, however, much more complex.

It appears likely, if not unavoidable, that Earth's collisional history had a multi-faceted influence on habitable conditions during the Hadean by affecting near-surface topography and geology, delivery and mobilization of key compounds, and formation of impact-generated hydrothermal systems. Current models of the Hadean suggest that high-standing continental crust may not have been prevalent (e.g., Bada & Korenaga, 2018); however, excavation and ejecta deposition by large impacts would have created sustained topography, including dry land and accompanying shallow water pools likely important to prebiotic chemistry (e.g, Ross & Deamer, 2016; Damer & Deamer, 2020). Deep sea hydrothermal events have been emphasized as a potential site of prebiotic compound synthesis and chemotropic life (e.g., Martin et al., 2008), and widespread impact-driven subaerial and subaqueous hydrothermal systems could have stimulated prebiotic chemistry pathways (e.g., Cockell, 2006). In addition, impacts supply heat and drive mixing in a planet's interior, resulting in the delivery and mobility of key biotic elements, including enhanced release of volatiles to the atmosphere from Earth's interior (e.g., CHNOPS; Marchi et al., 2016; Grewal et al., 2019). Other than delivery of essential elements and organic molecules that may have helped fuel life's beginnings, very little attention has been paid to the full range of consequences, both positive and negative, associated with large and frequent early impacts.

### 3.4 Our stellar neighborhood

Solar evolution models suggest the luminosity of the Sun was ~70% of current levels just after entering its main sequence hydrogen-burning phase 4.6 Ga (Bahcall et al., 2001). The juxtaposition of this low solar luminosity with the geological and geochemical evidence for warm (ice-free, liquid-water-rich) conditions on the Hadean and Archean Earth led to the so-called 'faint young Sun paradox' (e.g., Feulner, 2012). Recent 3D climate modeling has suggested that the combination of high $CO_2$ levels, cloud feedbacks, and modestly elevated $CH_4$ levels would resolve the apparent climate paradox (reviewed in Charnay et al., 2020). Warming of the early Hadean atmosphere may also have been enhanced by the existence of impact-generated/released reduced greenhouse gases and collisional-absorption complexes (Marchi et al., 2016; Zahnle et al., 2020).

While the early Sun's overall luminosity was lower, contemporary studies of young solar analogs have demonstrated that the early Sun's UV activity was significantly elevated compared to modern levels, with integrated high-energy emission >6 times larger than that of the modern Sun and more frequent coronal mass ejection (CME) events with attendant spikes in both XUV (X-ray and ultraviolet) and high-energy particle fluxes (Ribas et al., 2005, 2010). The multifold consequences of elevated solar XUV activity include substantially higher photodissociation rates of key molecules in the early atmosphere and enhanced surface UV fluxes relevant to synthesis of prebiotic compounds essential for the origin of life. Further, robust photolysis of $CO_2$, $H_2O$, and N-bearing species would have delivered to the early oceans a stream of possible electron donors and acceptors for simple chemosynthetic metabolism, including CO and $NO_x$ compounds (e.g., Kasting, 2014; Wong et al., 2017).

Differences in the solar NUV ($\lambda > 200$ nm) radiation would have been consequential for key prebiotic processes such as ribonucleotide and sugar synthesis pathways that rely on critical photochemical steps (e.g., Ranjan & Sasselov, 2016). Moreover, generation of energetic particles from CME events and subsequent chemical interactions with the early atmosphere may have profoundly influenced the delivery of organic compounds and maintenance of planetary climate (Airapetian et al., 2016, 2020). While progress abounds, we must continue to refine our understanding of the early Sun's connection with planetary processes in the atmosphere, oceans, and specifically as related to prebiotic chemistry on a young Earth.

### 3.5 Blue-sky research?

Earth's earliest atmosphere, rather than being blue, may have episodically been an orange haze like Titan's today due to photochemical reactions under impact-induced, methane-rich conditions (Zahnle et al., 2020) that may also have contributed to the abiotic production of life's building blocks. The controls and composition of the early atmosphere are still much debated, and solid Earth volatile emanations did not necessarily define the composition of the atmosphere (Trail et al., 2011). In some models, the atmosphere's composition is the source of key reactants needed to yield products required for prebiotic chemistry. These compositions also modulated surface temperature under subdued solar input (Charnay et al., 2020). Further, the presence and strength of the geomagnetic field would have regulated retention of an early atmosphere, and remanent magnetization associated with pre-4.0 Ga zircons suggests an early terrestrial magnetic field prior to 4.0 Ga. However, there is still no consensus on whether the magnetic carriers record primary information or later alteration (Tarduno et al., 2020; Borlina et al., 2020).

The earliest atmosphere is virtually unknown despite its critical role in maintaining clement temperatures and thus oceans at Earth's surface largely through a time-varying greenhouse balance of $CO_2$, methane, and water vapor (Zahnle et al., 2010). Further, the organic molecular products of photochemical reactions in the atmosphere may have been the feedstock that fueled prebiotic chemistry. The formation and evolution of the early atmosphere must be reconstructed in light of wide-ranging considerations: (i) an emerging geochemical database from zircons and the sparse rock record; (ii) models for early evolution of the Sun and related photochemistry; (iii) modeled tectonic/geodynamic, hydrothermal, and impact-related controls on degassing of deep-sourced volatiles; (iv) early differentiation of Earth's interior and related cooling more generally; and (v) weathering of seafloor and perhaps exposed landmasses, among other controls. This robust perspective should give us an exciting new environmental landscape for exploring prebiotic chemistry—while at the same illuminating critical factors for other, distant life-generating and life-sustaining systems.

### 3.6 What lies beneath?

The frustrations that have plagued many efforts to model prebiotic pathways to life's origins under low-temperature Earth-surface conditions have spurred intense consideration of alternatives, including deep-seafloor hydrothermal systems (Martin et al., 2008). Interest in hydrothermal systems is driven in part by the recent discovery and study of modern analogs, such as the Lost City vent field. There and at many other sites, serpentinization has become a hot topic because of the associated sources of energy and coupled pathways of organic synthesis and their possible broad relevance—extending from the origins of life on early Earth to present-day ocean worlds such as Enceladus and Europa. Such systems abound with chemical disequilibria, a condition that factors prominently in models for life's emergence, including extreme pH and redox gradients specific to alkaline hydrothermal systems (Russell et al., 2014). These settings also show strong potential for mineral catalysis of organic reactions through the presence of iron sulfides, for example—and experimental simulations (e.g., Barge & White, 2017) illuminate the range of related pathways by which organic compounds (that are stable under vent conditions) can be synthesized and accumulate from simple molecules. Hydrothermal vents provide a practical example of how local environmental conditions may have factored in prebiotic chemistry by constraining the chemical inventory and reaction pathways.

## 4.0 GRAND CHALLENGES AND RECOMMENDATIONS

The overarching goal must be to blend studies of prebiotic chemistry and related processes leading to the origins of life with a comprehensive, parallel, highly interdisciplinary deconstruction of Earth's earliest environments. Advances on this frontier are already showing promise thanks in part to the Prebiotic Chemistry and Early Earth Environments (PCE3) Research Coordination Network (RCN) within NASA's Planetary Science Division. The self-described nature of that group is an assembly of researchers *"striving to transform the origins of life community by breaking down language and ideological barriers and enhancing communication across the disciplinary divide between early Earth geoscientists and prebiotic chemists. We hope to cultivate a new paradigm across the community in which models for the emergence of a prebiotic pathway are rooted in realistic planetary conditions, and the dynamics and constraints of early Earth environments are fully integrated into origins hypotheses."*

Activities on this front must continue and grow in fundamental ways:
(i) by investing more comprehensively in studies of Earth's earliest environments prior to 4.0 billion years ago and incentivizing such research through its relevance to prebiotic chemistry.
(ii) by strongly encouraging related interdisciplinary projects and proposals bridging the environmental and prebiotic chemistry communities in unprecedented ways through existing and enhanced funding opportunities including those at large scale, such as the ICAR (Interdisciplinary Consortia for Astrobiology Research) program.
(iii) by working with greater consistency and impact with those focused explicitly on life detection beyond Earth—including much greater involvement in mission design, data analysis, and target selection for early Mars, ocean worlds, and exoplanets.

Searches for distant life must begin by asking whether life might have begun in those worlds during their earlier histories. We can reasonably and often reliably infer those early chapters now—through knowledge of stellar evolution, for example—but we must pursue this goal more rigorously and expansively through mission selection, related sampling/analyses, and subsequent data-based simulations. Most critically, though, this strategy will be compromised without a better understanding of the origins-environmental relationship on Earth. Restated, it is risky to search for life in any given system without asking if and how life might have started there, and any search engine for life beyond on planet must include an understanding of our own beginnings and specifically the relationships among Earth's earliest environments and the prebiotic chemical pathways that put life into motion.

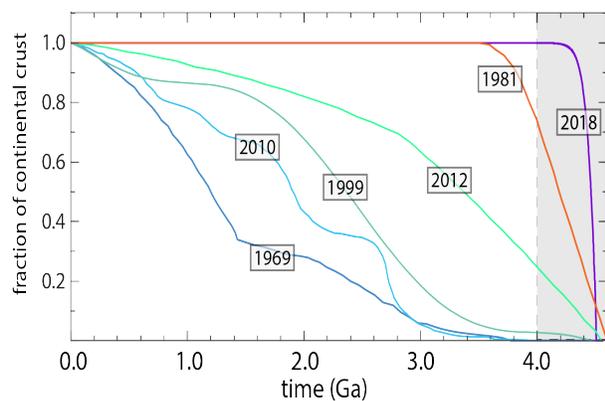 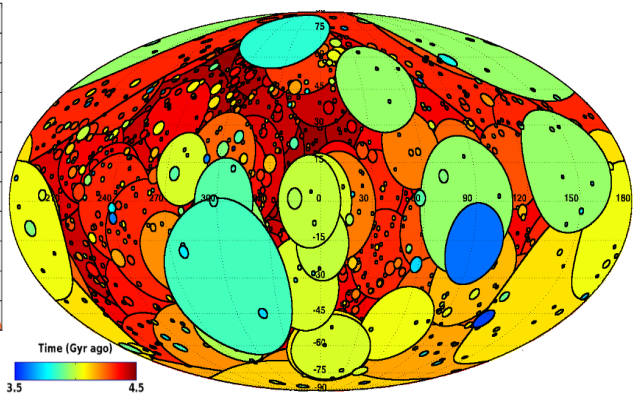

**Figure 1:** Representative published models for growth of continental crust normalized to present-day volume (after Korenaga, 2018). Note the large range of disagreement. Growth models are annotated with the year of publication.

**Figure 2:** A Mollweide projection of the cumulative record of terrestrial craters (3.5-4.5 Ga) in a characteristic Monte Carlo collisional simulation (after Marchi et al., 2014).